# How ChatGPT and Gemini View the Elements of Communication Competence of Large Language Models: A Pilot Study

Goran Bubaš

University of Zagreb Faculty of Organization and Informatics

Pavlinska 2, 42000 Varaždin

Croatia

`gbubas@foi.unizg.hr`

**Abstract**. A concise overview is provided of selected theoretical models of communication competence in the fields of linguistics, interpersonal communication, second language use, and human-robot interaction. The following practical research consisted of two case studies with the goals of investigating how advanced AI tools like ChatGPT and Gemini interpret elements of two communication competence theories in the context of Large Language Model (LLM) interactions with users. The focus was on these theoretical approaches: (1) an integrated linguistic-interpersonal model and (2) an interpersonal 'human-humanoid' interaction model. The conclusion is that both approaches are suitable for a better understanding of LLM-user interaction.

## 1 Introduction

According to Chang et al. (2024), because of their widespread use, the evaluation of Large Language Models (LLMs) from various perspectives has become crucial from the *task execution* level to the *social* level. Chang et al. state that this incorporates the capacity of LLMs for *natural language understanding and generation*, including *fluency* (the ability to produce textual content "that flows smoothly, maintaining a consistent tone and style") and *human alignment* (the degree to which the output relates to "human values, preferences, and expectations", as well as "societal norms and user expectations, promoting a positive interaction with human users"). In the theoretical part of the following study, first, a brief overview of the development of *linguistic* approaches to human communication, as well as approaches to communication competence from an *interpersonal communication skills* perspective, is presented. Following this, there are concise summaries of (a) the *integrated model of linguistic and interpersonal communication competence in second language use* developed by Bubaš and Kovačić (2019) and of (b) an early model of *communication competence of artificial systems* introduced by Bubaš and Lovrenčić (2002). In the two case studies that are also presented in this paper, the communicative abilities of LLMs are investigated in detail from these two theoretical perspectives. It must be noted that, for this examination, the 'views' of advanced LLMs of the ChatGPT and Gemini family are utilized regarding the applicability of the content of the two theoretical studies (Bubaš and Kovačić, 2019; Bubaš and Lovrenčić, 2002) for interpreting various communication ('competence') aspects of interaction of LLMs with their users. The presented study is relevant because communication competence related aspects of LLM-user interaction have been scarcely and only fragmentarily investigated. Therefore, this study represents a novel theoretical and practical approach to the investigation of communication of LLMs with their users. In the continuation of this article, a theoretical introduction is provided with two theoretical models of communication competence explained in more detail. After the definition of the main goal and research questions, study methodology is explained with details of prompting procedures for two case studies. The results of the two case studies are interpreted in relation to the research questions (RQ1 – RQ3) and briefly discussed. The article resumes with a concise conclusion and limitations of the study.

## 2 Communication Competence Theoretical Models

This theoretical section first provides a summary of the introduction of the concept of *communication competence* in the fields of *linguistics* (in *subchapter 2.1*) and *communication science* (in *subchapter 2.2*) with a specific focus of the latter on *interpersonal interaction*. Then, the model of linguistic and interpersonal communication competence in second language use is briefly outlined (in *subchapter 2.3*) that *integrates linguistic and interpersonal aspects of communication competence* and can be used in theoretical interpretations of communication of LLMs with users. Finally, a concise summary of the *model of communication competence of artificial systems* is presented (in *subchapter 2.4*) that was created primarily for the *theoretical interpretation of* *human-robot interaction*, but can also be adapted to the context of interaction of LLMs with users.

## 2.1 Linguistics and communication competence

The concept of *linguistic competence* was developed by Chomsky (1965, pp. 3-10) as an 'idealized' *intrinsic knowledge* of language possessed by a speaker-hearer, distinct from the actual, *real-world language performance* of the individual. This limited view of linguistic competence was challenged in a series of publications by Hymes (for instance, see Hymes, 1972), who introduced the term *communicative competence*. For Hymes, this concept included not only linguistic competence but also *pragmatic*, *socio-cultural*, and *sociolinguistic* aspects of language use.

Within the linguistic line of research, Canale and Swain (1980) proposed the following components of communication competence: *grammatical* (knowledge of grammar, syntax, vocabulary, morphology, and phonology); *sociolinguistic* (how to appropriately use language in a social context, including situational and cultural aspects); *strategic* (use of verbal and non-verbal strategies concerning grammatical and sociolinguistic competence). Canale (1983) later added the component of *discourse* competence as the capacity to "combine grammatical forms and meanings to achieve a unified spoken or written text in different genres".

The notable theoretical contributions in this field include the work of Bachman and Palmer (1995, 67-75), who proposed the term *communicative language ability*, one component of which is *language knowledge*, which can be further divided into (a) *organizational knowledge*, composed of grammatical and textual knowledge, and (b) *pragmatic knowledge*, consisting of functional and sociolinguistic knowledge. According to Bachman and Palmer, *functional knowledge* includes the expression of the individual's experiences with the real world, manipulation of other people, and development of relationships with them. Also, *sociolinguistic knowledge* enables the adaptability of created or interpreted language to a specific interactional context, for instance, in terms of applicable conventions and potential appropriateness evaluations. Finally, these authors consider *strategic competence* as a set of metacognitive components operating in the practical areas of goal and task setting, assessment of the task and its actualization, as well as planning for completing or implementing the goal or task.

Various versions of linguistic models of communication competence have also been developed, but most of them encompass similar constructs as previously mentioned. For instance, the model by Celce-Murcia et al. (1995) introduces a triangular organization of *sociocultural*, *linguistic*, and *actional* competence, which are all in interaction with *discourse* and *strategic* competence.

There are several historical overviews on the approaches to communication competence in linguistics and language learning (including, for instance, Bagarić & Mihaljević Djigunović, 2007), as well as efforts to integrate the related theoretical concepts (e.g. Usó-Juan & Martínez-Flor, 2006) for application in language learning. It is important to mention that, while the domain of *language teaching and learning practice* has predominantly focused on the *four skills* (listening, speaking, reading, and writing) as learners' abilities for effective language use, the field of *linguistics* (and applied linguistics in particular) has seen a continuation of the interest in the concepts of *grammatical* competence and language knowledge, as well as the variations of *sociolinguistic*, *strategic*, and *discourse* competence, including their *pragmatic aspects* and integration into more holistic models.

Recently, efforts have been made to associate linguistic aspects of communication competence with *conversational artificial intelligence* and *Large Language Models* like ChatGPT.

## 2.2 Interpersonal communication and communication competence

In the field of interpersonal communication, as a discipline within *Communication Science*, the most extensive early theoretical research on communication competence was performed by Spitzberg and Cupach (1984; 1989). Their work aimed to conceptualize communication competence in *interpersonal social interaction*, distinct from the theoretical and practical areas of linguistics. These authors investigated and reflected on numerous studies of interpersonal communication skills to form a general approach with a tripartite model comprising (1) interpersonal or social *skills* that may contribute to communication competence, (2) *knowledge* that can be, for instance, procedural or tacit/indeliberate, and (3) *motivation* that may be broadly denoted by the approach-avoidance dimension in interactions. Spitzberg and Cupach (2002) also emphasized the need for criteria of skillful interpersonal interaction, like, for instance, *effectiveness*, *efficiency*, *appropriateness*, and *satisfaction*. In their "heuristic for taxonomies" of more than 100 labels for communication skills found in the literature, they also defined the "macroscopic" level of interpersonal interaction, as well as "microscopic" and "mezzoscopic" skills. For the 'in-depth' perspective of interpersonal communication competence, apart from the condensed overview of the historic developments around this term in the communication discipline by Backlund and Morreale (2015), it is also useful to consider the analysis of communicative competence as a theoretical concept by Wilson and Sabee (2003). It must be emphasized that a newer model of *computer-mediated communication (CMC) competence*, developed by Spitzberg (2006), (a) is based on the concepts of knowledge, motivation, and skills, (b) includes outcomes like appropriateness, effectiveness, coorientation, satisfaction, and relational development, and (c) also comprises several additional components including various media, message, and contextual factors.

It can also be mentioned that an effort to utilize empirical research on the dimensions of *human* interpersonal communication competence (see: Bubaš et al., 1999; Bubaš, 2001a; Bubaš, 2001b; Bubaš, 2003) with potential communication competence dimensions of *artificial behavioral (intelligent) systems that interact with humans* was performed more than two decades ago by Bubaš and Lovrenčić (2002).

Regarding LLMs like ChatGPT, more recently, researchers have focused on a small number of specific interpersonal communication skills that may be manifested by this AI tool, including *empathy* (see: Sorin et al., 2024; Luo et al., 2024; Welivita & Pu, 2024) and *persuasion* (Bai et al., 2025; Rogiers et al., 2024; Goel et al., 2024), rather than on a larger set of skills and their manifestation in complex behavior similar to *interpersonal communication competence* in general. By contrast, Bubaš (2024) has demonstrated the ability of LLMs like ChatGPT 4 and ChatGPT 4o, Copilot, Claude 3.5 Sonnet, and Gemini 1.5 Pro, to analyze existing items and develop new ones for self-assessment scales related to a large number of different interpersonal communication skills.

## 2.3 Integrated model of linguistic and interpersonal communication competence in second language use

To integrate (a) the *linguistic* elements with (b) the *interpersonal* interaction and *skills* elements of theoretical models of communication competence, Bubaš and Kovačić (2019) have developed a *conceptual model of communication competence in foreign language* (L2) *use* (CMCC-L2) and interpreted this model in the context of teaching and learning L2 and *English as a Foreign Language* (EFL). This new model adapted and integrated the "framework of communicative competence integrating the four skills" (Usó-Juan & Martínez-Flor, 2006), the molar and molecular view of interpersonal communication competence (Spitzberg & Cupach 1984), different levels of observing interaction processes – microscopic, mezzoscopic, macroscopic (Spitzberg, 1994), including similar levels of analysis in the social ecological framework (Bronfenbrenner, 1977; Ting-Toomey, 2012), and communication accommodation theory (Zhang, 2018). The outline of the levels of the *conceptual model of communication competence in foreign language use* (L2) is presented in Table 1 (mostly adapted from Usó-Juan & Martínez-Flor, 2006).

**Table 1.** Levels in the conceptual model of communication competence in a foreign language (L2)

| LEVEL | COMPETENCE TYPE |
|---|---|
| Supra | **Social/intercultural** *(utilization of knowledge, social/cultural cues, and skills to understand the interaction environment and appropriately perform wide-ranging sequences of intentional communication acts)* |
| Macro | **Strategic/adaptive** *(knowledge and use of interaction strategies, learning to adapt and advance in competence, utilization of specific skills to enhance ability and overcome barriers)* |
| Mezzo | **Pragmatic/action/discourse** *(performing and interpreting concise speech acts, monologue, and dialogue according to participant and situational variables)* |
| Micro | **Linguistic** *(lexicon, phonology, orthography, morphology, syntax, sentence sequencing)* |

In Fig. 1, using the basic *motivation-knowledge-skills* concept as the potential for interpersonal competence, as well as *outcomes* concepts of Spitzberg (2006) and Spitzberg and Cupach (2002), the *conceptual model of communication competence in foreign language (L2) use* (CMCC-L2) is presented, adapting and further developing these models to define "enactment of skill(s)" and "L2 competence confirmation" (or "social outcomes"). The list of "skills" as a potential for competence is comprehensive: *willingness to communicate, initiation of interaction, listening, self-disclosure, nonverbal sensitivity, self-monitoring, impression management, questioning, empathy, persuasion/assertiveness, conversational skill, expressiveness, social support, and interaction management*. The "enactment of skill(s)" is at the interpersonal interaction levels of (1) *pragmatic/action/discourse*, (2) *strategic/adaptive*, and (3) *social/intercultural* competence and includes selected skills for each of the previous levels (1-3). Finally, the "L2 competence confirmation" and related "social outcomes" can be: *understanding, appropriateness, influence, coordination, satisfaction, cooperation, efficacy (goal attainment), attractiveness, relationship, inclusion, socialization, learning, competence enhancement, self-development, growth, well-being, and community building*. Basically, this model illustrates how specific communication *skills* (as *potential* or *capacity*, including *knowledge* and *motivation* components) can be used to facilitate or enhance the practical manifestation of the *dimensions of L2 communication competence* (enactment of skills) and produce various social outcomes that are confirmation of L2 communication competence.

Some theoretical concepts from the field of linguistics (including several elements that are mentioned in the *conceptual model of communication competence in L2 use* by Bubaš and Kovačić, 2019), have already been utilized in the analyses of communicative abilities of LLMs like ChatGPT, however a similar *integrative model* has not yet been found in the literature. For instance, an extensive survey of language model behaviors analyzed *separately* their abilities regarding *syntax*, *semantics* and *pragmatics* (Chang & Bergen, 2024), while more focused studies, among others, for instance, investigated linguistic competence of ChatGPT from the aspects of *pragmatic awareness* regarding impoliteness (Andersson & McIntyre, 2025), expressive and receptive *pragmatic skills* (Barattieri di San Pietro, 2023), *sociolinguistic competence* (Duncan, 2024), formal versus functional *linguistic competence* (Mahowald et al., 2024), *linguistic competence* regarding Italian dialects (Lilli, 2023), and *intercultural competence* (Lee, 2025).

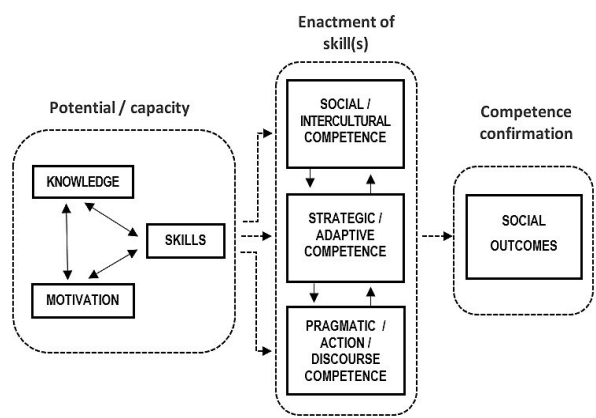

**Figure 1.** Relationship among the elements of the *conceptual model of communication competence in foreign language (L2) use* (adapted from Bubaš & Kovačić, 2019)

## 2.4 An early model of communication competence of artificial systems

Four decades ago, Spitzberg & Cupach (1984) identified more than 100 labels for communication skills and traits in scholarly literature in relation to human interpersonal interaction. Much later, using self-assessment scales for 23 different communication skills and traits and factor analyses, in empirical investigations of the *dimensions* of human interpersonal communication competence (Bubaš et al., 1999; Bubaš, 2001a; Bubaš, 2001b; Bubaš, 2003) the following six potential dimensions were identified (with numerous specific communication skills as their subcomponents): (1) encoding and decoding, (2) intentionality, (3) communication effectiveness, (4) other-orientedness, (5) expressiveness, and (6) social relaxation. These empirically revealed dimensions were used to develop a conceptual model of *communication competence of artificial systems* (CCAS) that interact with humans (Bubaš & Lovrenčić, 2002). However, at the time this CCAS model was developed (in the year 2002), when compared to the dimensions of *human* communication competence, the developments in *humanoid robotics* were only at the general level of the *decoding and encoding dimension* and the rudimentary *expressivity dimension*. It must be emphasized that in the CCAS model, a formalization of the generalized dimensions of interpersonal communication competence was outlined (in the context of *human-humanoid interaction*), including skills and traits that contribute to those dimensions. Finally, in their paper, Bubaš and Lovrenčić (2002) have suggested possible implementations of the formalized CCAS model in the design of interfaces that could enhance the *effectiveness* and *appropriateness* in communication of an artificial (intelligent) system with humans. It was only after the launch of OpenAI's ChatGPT in November 2022 that most of the other components of the CCAS model could be applied to ChatGPT and other families of LLMs (e.g., Anthropic's Claude, Google AI's Gemini, etc.) to interpret interactions between LLMs and their users.

In the article about the CCAS model, the intention was to present in detail the following dimensions (and not more than *four related communication skills* for each dimension): decoding and encoding (*nonverbal sensitivity, verbal understanding, verbal and nonverbal encoding, self-monitoring*), communication effectiveness (*initiation of interaction, assertiveness, interaction management, adaptability*), and other-orientedness (*empathy, support, self-disclosure, collaboration*). All six dimensions of CCAS are depicted in Fig. 2.

In addition, the CCAS model organized communication behaviors of artificial systems in the following levels:
- *Macroscopic* level (dimensions of communication competence).
- *Mezzoscopic* level (sets of skills and traits that are related to each of the dimensions of communication competence).
- *Microscopic* level (sets of "communication primitives"; moderately complex verbal and nonverbal behaviors, behavioral scripts, skill-related knowledge, communication traits, interaction tactics, etc., that contribute to the pragmatic enactment of a specific communication skill).
- *Atomic* level (sets of simple perceptual or motoric behaviors that are engaged in the realization of the specific communication "primitives").

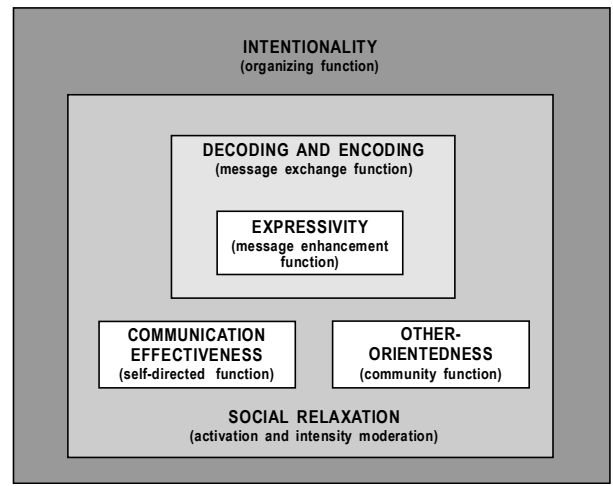

**Figure 2.** The dimensions of communication competence of artificial (intelligence) systems (source: Bubaš & Lovrenčić, 2002).

## 3 Goals and research questions

The *main goals* of this preliminary study were to investigate how advanced LLMs like ChatGPT o3, ChatGPT o4-mini, ChatGPT 4.5, Gemini 2.5 Flash and Gemini 2.5 Pro view the elements of the communication competence of LLMs in their interaction with users according to the (a) more linguistic approach of the *conceptual model of communication competence in foreign language (L2) use* (CMCC-L2), and (b) more interpersonal approach of the model of *communication competence of artificial systems* (CCAS). These two models were chosen based on the previously presented overview of theoretical literature as some of the most relevant, elaborate communication competence models that may be applied to LLM-user interaction and they also included listings of related communication skills. In comparison to similar studies, the goals of this study were focused on the 'self-evaluation' of LLMs' own interaction with users, instead of those performed by users or experts.

According to the previously listed main goals of this pilot study, the following research questions were defined:

RQ1: Can advanced ChatGPT and Gemini LLMs understand the basic elements of the CMCC-L2 and CCAS theoretical models?

RQ2: Can advanced ChatGPT and Gemini LLMs use the elements of CMCC-L2 and CCAS models to interpret their knowledge of the way LLMs interact with their users?

RQ3: How useful are the CMCC-L2 and CCAS models for eliciting information from the advanced ChatGPT and Gemini LLMs about the communication skills of LLMs in their interaction with users?

# 4 Methodology

The methods that were used in this pilot study can be described as *elicitation of knowledge from Large Language Models using a specific theory-driven design of prompts*. The qualitative methodology was characterized by *interpretative analyses* that included *two case studies*, each using one of the two previously presented theoretical frameworks: CMCC-L2 and CCAS. In *comparative analyses,* the CMCC-L2 and CCAS models were examined by *three* advanced ChatGPT models and *two* of the latest Gemini models (in early July 2025) regarding the *context of interaction of LLMs with users*. In essence, a *set of prompts* was designed in each case study for ChatGPT and Gemini to *elicit their responses* that reflect the components of those communication competence models (CMCC-L2 and CCAS) and determine the degree of their alignment with the LLMs' outputs, *as 'viewed' by ChatGPT and Gemini*. This provided insights into the competence models' interpretative potential and actual 'interpretations' of ChatGPT and Gemini regarding the interaction of LLMs with the users. In other words, the two case studies were designed to provide insight into the ways ChatGPT and Gemini 'explain' the elements of 'communication competence' of LLMs within the framework of CMCC-L2 and CCAS. Therefore, after the previous work of Bubaš et al. (2024) on *usability* and *user experience* aspects of *Human-Computer Interaction* (HCI) with LLMs, the current study is positioned more in the overlapping areas of *Human-AI Interaction* (HAI) and *Explainable AI* (XAI).

Since the prompting of ChatGPT and Gemini regarding the CMCC-L2 and CCAS models was performed as two separate case studies, the methodology for each differed slightly and will be explained separately.

The practical part of the methodology enables reproducibility since other researchers can use the comparable LLMs, published papers (Bubaš & Kovačić, 2019; Bubaš & Lovrenčić, 2002), and prompts for equivalent studies. However, exactly the same results should not be expected in replication studies, since the output of LLMs is inherently stochastic, even when using the same model (e.g., Gemini 2.5 Pro) with precisely the same prompt wording, especially when prompt and response complexity are high or when the prompt is run in a new chat session. Also, with the development of newer versions of ChatGPT and Gemini LLMs, the older versions (i.e., those that were *state-of-the-art* at the time this study was conducted) may become unavailable for similar studies. It must be emphasized that replicability was not tested with Grok and DeepSeek models.

## 4.1 Instruments

The instruments that were used in this study were OpenAI's *ChatGPT* advanced reasoning models o3, o4-mini, and GPT 4.5, as well as Google DeepMind's *Gemini* 2.5 Flash and 2.5 Pro models. The LLMs were used with the *ChatGPT Plus* and *Google AI Pro* subscription plans. LLM's options like 'Web search', 'Deep research' and the 'temperature' setting were not used in prompting.

The researcher also used his and his coauthors' published and publicly available articles (Bubaš & Kovačić, 2019; Bubaš & Lovrenčić, 2002) in PDF format, along with JPEG images of figures from those articles on the CMCC-L2 and CCAS models, which were also considered as 'instruments' in this study after uploading them to ChatGPT and Gemini LLMs during prompting.

There were no human subjects in the two presented case studies, and the prompting of ChatGPT and Gemini was not in its form different from everyday use of those tools by researchers, faculty, and students. For both case studies, the procedure and prompts were first tested and refined with the *ChatGPT 4o* model. To ensure that the images in the PDFs of the articles are processed by ChatGPT and Gemini tools, the images in JPEG format were uploaded, and the LLM was asked to 'read' and reproduce the content of the image. It was important to determine that the text in the image was 'read' and the graphic content 'understood' by the LLMs, as well as that the ChatGPT and Gemini tools could interpret the images in the context of the full article in the PDF format that was also uploaded for further interpretation of the elements of the two theoretical models of communication competence. The previously described procedure was performed with all ChatGPT (o3, o4-mini, ChatGPT 4.5) and Gemini (2.5 Flash, 2.5 Pro) tools separately for the article on CMCC-L2 (Bubaš & Kovačić, 2019) in the *first* case study, and for the article on CCAS (Bubaš & Lovrenčić, 2002) in the *second* case study.

## 4.2 Procedure of the *first* case study

The *first case study* used the advanced ChatGPT and Gemini tools in combination with the PDF of the article on CMCC-L2 (Bubaš & Kovačić, 2019).

The procedure for using ChatGPT and Gemini tools in the first case study was performed in the following steps:

i. Fig. 2 from the article was uploaded in the LLM in JPEG format with the following prompt (#1): *List the skills in each segment of the model shown in the image, and provide a brief explanation of each segment.* The ability of the LLM to read and interpret the content of the image was verified before proceeding to the next step.

ii. Fig. 1 from the article was uploaded to the LLM in JPEG format with the following prompt (#2): *In the image, what are the descriptions and levels of the following: (1) linguistic competence, (2) pragmatic/action/discourse competence, (3) strategic/adaptive competence, and (4) social/intercultural competence?* Again, the ability of the LLM to read and interpret the content of the image was verified before proceeding further.

iii. The PDF document with the article (Bubaš & Kovačić, 2019) was uploaded to the LLM. To evaluate the ability of the LLM to interpret the theoretical content of the article, a prompt (#3) with the following explanation and instruction was provided to the LLM: *The two images you previously described are from the PDF document (Fig. 1 and Fig. 2) in the scholarly article entitled 'Communication Competence Related Skills in the Context of Student Performance and Teaching in the EFL Classroom.' Keeping in mind the content of the images (Fig. 1 and Fig. 2) and the text in the PDF document, explain the conceptual model of communication competence in a foreign language (L2), including how it incorporates interpersonal communication skills.* The ability of the LLM to explain the CMCC-L2 model was evaluated by the first author of that article before checking the capacity of the LLM to contextualize this model to LLM communication with users.

iv. The 'awareness' of the LLM of the possibility to contextualize the CMCC-L2 was checked with the following prompt (#4): *If the context of second language use and teaching is excluded, the model you described can be used to interpret and explain the interaction between Large Language Models (LLMs) like ChatGPT and Gemini and their users. Theoretically, it could be reformulated as a 'conceptual model of communication competence of Large Language Models (LLMs).' Do you understand this possibility?* After validating the confirmation by the LLM, the following final prompt was used in the prompting sequence that was the same for all LLMs.

v. The ability of the LLMs to associate specific communication skills that were mentioned in the article and CMCC-L2 (including the related LLMs knowledge and interpretative capacity) was 'elicited' with the following prompt (#5): *Try to connect as many communication skills from the two original images (Fig. 1 and Fig. 2 from the PDF document) and the text of the article with the interaction of LLMs and their users. Provide 1–2 examples for each specified communication skill, showing how it is associated with a specific type and form of interaction between an LLM and a user. Use the names/labels of communication skills that appear in the PDF document (article) as much as possible.*

vi. The responses to the last two aforementioned prompts, collected from selected ChatGPT (o3, o4-mini, ChatGPT 4.5) and Gemini (Flash 2.5, Pro 2.5) tools, were compared in the context of 'communication competence related elements of LLMs' communication with users. Specific additional prompting was used to elicit further information on this topic to complete the initial/pilot investigation for the first case study.

### 4.3 Procedure of the *second* case study

The *second case study* used a comparable methodology to the first case study. The difference was that the advanced ChatGPT and Gemini tools were used with the PDF of the article on the CCAS model (Bubaš & Lovrenčić, 2002).

The main prompting procedure was analogous and therefore only the prompts that were used will be listed (in italics):

i. *In the PDF document is an article entitled 'Implications of Interpersonal Communication Competence Research on the Design of Artificial Behavioral Systems that Interact with Humans.' Explain the conceptual model of communication competence for artificial (intelligent) behavioral systems presented in the article, including how it incorporates interpersonal communication skills.* (prompt #1)

ii. *In the image, what are the descriptions of the following dimensions of CCAS: (1) intentionality, (2) social relaxation, (3) decoding and encoding, (4) expressivity, (5) communication effectiveness, and (6) other-orientedness? How would you explain the organization (i.e., the relationships) among these dimensions as depicted in the image?* (prompt #2; see Fig. 2 in this article)

iii. *The model you described from the PDF document and the image (Fig. 2) can be used to interpret and explain the interaction between Large Language Models (LLMs) like ChatGPT and Gemini and their users. Theoretically, it could be reformulated as a 'conceptual model of communication competence of Large Language Models (LLMs).' Do you understand this possibility?* (prompt #3)

iv. *Try to connect as many communication skills from the PDF document (i.e., the article text) as possible with interactions between large language models (LLMs) and their users. For each specified communication skill, provide 1–2 examples that illustrate how it relates to a specific type or form of interaction between an LLM and a user. Use the exact names or labels of communication skills as they appear in the PDF document whenever possible.* (prompt #4)

v. *Using the text of the article titled 'Implications of Interpersonal Communication Competence Research on the Design of Artificial Behavioral Systems that Interact with Humans' and the image (Fig. 2), describe the dimensions of CCAS in the context of interactions between large language models (LLMs) and users: (1) intentionality, (2) social relaxation, (3) decoding and encoding, (4) expressivity, (5) communication effectiveness, and (6) other-orientedness. At the end, write 1–2 sentences estimating how applicable the CCAS model is to the context of LLM–user interaction, and whether it could be reformulated as a 'conceptual model of communication competence for Large Language Models (LLMs).'* (prompt #5)

## 5 Results and discussion

The results of the two case studies will be interpreted separately concerning the research questions (RQ1, RQ2, and RQ3).

### 5.1 Results of the first case study

Considering the first research question (RQ1) – *Can advanced ChatGPT and Gemini LLMs understand the basic elements of the CMCC-L2 and CCAS theoretical models?* – All advanced ChatGPT (o3, o4-mini, ChatGPT 4.5) and Gemini (2.5 Flash, 2.5 Pro) tools were able to 'read' and 'interpret' the content of the JPEG image of Fig. 1 and Fig. 2, as well as the PDF document with the article *'Communication Competence Related Skills in the Context of Student Performance and Teaching in the EFL Classroom'* and its explanation of CMCC-L2 (prompts #1, #2, and #3). Furthermore, these ChatGPT and Gemini tools confirmed that they understood the possibility of using CMCC-L2 to explain the interaction between LLMs and their users (prompt #4), and also provided meaningful interpretation of the theoretical elements and selected communication skills in CMCC-L2 in the context of interaction between an LLM and users (prompt #5).

Regarding the second research question (RQ2) – *Can advanced ChatGPT and Gemini LLMs use the elements of CMCC-L2 and CCAS models to interpret their knowledge of the way LLMs interact with their users?* – Due to the limited length of this article, only the selected interpretations will be presented. In Table 2, the interpretations by advanced ChatGPT and Gemini tools of the interaction of LLMs with their users in the context of *CMCC-L2* and *CCAS* are illustrated. The process of selecting illustrative outputs from the LLMs in Table 2 and in subsequent tables of this article was based on: (a) the informational value of the LLMs' output, (b) the intention to demonstrate that all examined LLMs were able to produce valuable output, and (c) the tendency to avoid repetition of similar output by different types of LLMs.

**Table 2.** Selected interpretations from ChatGPT and Gemini tools regarding elements of LLMs' *linguistic competence*, *discourse / pragmatic / action competence*, *strategic / adaptive competence*, and *social / intercultural competence* (LLM outputs are shown in quotation marks)

---

**Linguistic competence** *(micro level)*
- „LLM Analog: its mastery of sub-word and word tokens, grammar, spelling, and the statistical patterns of sentence structure in its training data." ; „What this gives the model: the raw ability to produce well-formed strings of text in the target language." *[ChatGPT o4-mini]*
- „In LLMs: Refers to the ability of the model to produce grammatically correct, coherent, contextually appropriate, and semantically meaningful outputs. It involves lexical choice, syntax structuring, accurate language representation, and the foundational capacity to generate linguistically precise responses." *[ChatGPT 4.5]*

**Discourse / pragmatic / action competence** *(mezzo level)*
- "In LLMs: Refers to the model's proficiency in managing and adapting interactions based on different communicative situations and user expectations. For example, strategically clarifying ambiguous requests, gracefully handling user misunderstandings, effectively managing context shifts, adapting tone or style of responses, and proactively offering useful follow-up suggestions." *[ChatGPT 4.5]*
- "Concept: This level focuses on the LLM's ability to perform and interpret 'speech acts' (e.g., making requests, giving commands, providing information), and to engage in coherent monologues and dialogues by considering the user and situational variables. It moves beyond individual sentences to how language is used in context for specific communicative functions." *[Gemini 2.5 Flash]*

**Strategic / adaptive competence** *(macro level)*
- LLM operation: „The policy-and-planning layer: monitors ambiguity, requests clarification, invokes tools, follows safety rules, or changes strategy mid-conversation."; Example regarding clarification strategy: „Prompt 'Tell me about Apple' → model answers 'Do you mean the company, the fruit, or the record label?'— choosing to disambiguate before continuing." *[ChatGPT o3]*
- „LLM Analog: its use of higher-level strategies—asking clarifying questions when prompts are ambiguous, self-correcting hallucinations on the fly, adapting style and length to the user's feedback, using external tools or memory APIs when needed."; „What it accomplishes: overcoming 'breakdowns' (e.g. vague prompts), guiding the user back on track, and ensuring the exchange stays productive." *[ChatGPT o4-mini]*

**Social / intercultural competence** *(supra level)*
- LLM operation: „The alignment layer: modulates tone, mitigates bias, observes cultural taboos, expresses empathy, and sustains trust over extended interactions."; Example regarding bias mitigation & inclusivity: „When asked for 'a joke about lawyers,' the model offers a mild, non-offensive joke or gently refuses a sexist stereotype request—demonstrating ethical and culturally sensitive filtering." *[ChatGPT o3]*
- "This is the highest level, where the LLM uses its knowledge and skills to understand the broader context of an interaction and behave appropriately according to social and ethical norms." *[Gemini 2.5 Pro]*

---

The interpretations by ChatGPT and Gemini of **'linguistic competence'** of LLMs in Table 2 include the use of grammar and spelling rules, syntax and sentence structure, proper lexical choice, producing strings of text in a specific language, and coherent, semantically meaningful outputs. Various definitions by ChatGPT and Gemini of **'discourse / pragmatic / action competence'** of LLMs in Table 2 are related to management of turn-taking and context windows, keeping topics coherent, selection of an adequate speech-act format, switching registers, adaptation of interactions to specific communication situations and user expectations, and using language in context for specific functions. The aspects of **'strategic / adaptive competence'** of LLMs in Table 2, 'viewed' by ChatGPT and Gemini, include the use of diverse strategies, management of and adaptation to interactions, adjustment of responses, and overcoming barriers. Regarding the highest level of **'social / intercultural competence'** of LLMs, as 'perceived' by ChatGPT and Gemini and depicted in Table 2, it incorporates a higher level of awareness of social contexts, cultural specificities, social dynamics, ethical norms, and related user expectations, as well as the appropriate adaptation of tone, register, and output content, among others, to achieve desired outcomes.

Since CMCC-L2 also includes communication skills, the ChatGPT and Gemini tools were further asked (prompt #4) to interpret the communication skills mentioned in the Bubaš and Kovačić (2019) article in association with the interaction of LLMs with their users. Their representative responses for selected communication skills are displayed in Table 3.

A detailed inspection of the responses of ChatGPT and Gemini tools in Table 2 reveals how they 'interpret' the interaction of LLMs with their users by associating specific types of interaction with their semantic interpretation of a large set of communication skills. It must be noted that Bubaš (2024) has demonstrated that even the earlier versions of popular LLMs (ChatGPT 4, ChatGPT 4o, Microsoft Copilot, Claude 3.5 Sonnet, and Gemini 1.5 Pro) had a very high level knowledge of various interpersonal communication skills and the ability to generate items for related self-assessment scales having in mind different contexts (e.g. *face-to-face* versus *online* interaction).

Generally speaking, despite the potential for various interpretative biases of LLMs (e.g., cultural, positivity, and framing biases), the data presented in Table 2 and Table 3 contribute to a positive response to the third research question (RQ3) – *How useful are the CMCC-L2 and CCAS models for eliciting information from the advanced ChatGPT and Gemini LLMs about the communication skills of LLMs in their interaction with users?* Therefore, the methodology that was employed effectively 'elicited' the views of those AI tools regarding the communicative interaction of LLMs with users. Here are two sample quotes that are associated with RQ3 from the viewpoint of ChatGPT and Gemini tools:
- ✓ "Each of these parallels shows that LLMs don't just generate text—they enact a suite of interpersonal 'skills' that mirror those of human communicators, from listening and questioning, to empathy and strategic self-monitoring, thereby building a richly interactive, socially meaningful exchange." *[ChatGPT o4-mini]*
- ✓ "Each communication skill from the original model can thus meaningfully translate into real-world interactions between LLMs and users, offering structured guidance on improving communication effectiveness, enhancing user satisfaction, and facilitating productive outcomes in human-LLM interactions." *[ChatGPT 4.5]*

**Table 3.** Interpretations from ChatGPT and Gemini tools of *selected communication skills from CMCC-L2* in the context of the interaction of LLMs with users (the output of LLMs is in quotation marks)

| SELECTED COMMUNICATION SKILLS IN THE LLM-USER INTERACTION CONTEXTS |
|---|
| **Listening** *("textual receptive accuracy")* |
| • "The LLM parses the entirety of a user's prompt, keeps referents straight, and mirrors key words back to signal understanding." *[ChatGPT o3]* |
| • "A user provides a long, multi-part prompt. The LLM demonstrates 'listening' by addressing all parts of the prompt in its response, rather than just the last sentence." *[Gemini 2.5 Flash]* |
| **Self-monitoring** |
| • "Internal safety/policy checks, token limits, hallucination detectors." *[ChatGPT o3]* |
| • "Automatically evaluating its own responses for potential inaccuracies or bias and proactively indicating uncertainty, e.g., 'I'm not completely certain about this fact; would you like me to verify it further?'" *[ChatGPT 4.5]* |
| **Impression management** |
| • "Always maintaining a polite and professional tone, even when faced with aggressive or provocative user inputs." *[ChatGPT 4.5]* |
| • "A user asks for professional advice. The LLM responds in a formal, respectful, and authoritative tone, managing the impression it conveys to align with the user's expectation of professional discourse." *[Gemini 2.5 Flash]* |
| **Questioning** |
| • "Asks clarifying questions when the user goal is ambiguous or underspecified." *[ChatGPT o3]* |
| • "To assist in a business brainstorming session, the LLM asks a series of structured questions like 'What is the primary goal of this project?' and 'Who is the target audience?' to help the user flesh out their ideas." *[Gemini 2.5 Pro]* |
| **Empathy** |
| • "Expresses understanding of feelings, uses supportive language." *[ChatGPT o3]* |
| • "A user describes a difficult personal situation. The LLM responds in a supportive and non-judgmental tone, providing helpful resources or simply acknowledging the user's feelings." *[Gemini 2.5 Pro]* |
| **Persuasion/assertiveness** |
| • "Suggesting healthier habits ('A short walk every hour can boost focus—give it a try!') and providing compelling reasons." *[ChatGPT o4-mini]* |
| • "Assertively correcting misinformation without alienating users, e.g., 'That's a common misconception, but recent evidence suggests otherwise.'" *[ChatGPT 4.5]* |
| **Conversational skill** |
| • "Keeps turn-taking smooth, uses discourse markers, stays on topic, provides summaries." *[ChatGPT o3]* |
| • "A user begins a conversation about travel plans. The LLM maintains the topic, asks follow-up questions about destination or preferences, and provides relevant suggestions, demonstrating its conversational skill in a sustained dialogue." *[Gemini 2.5 Flash]* |
| **Expressiveness** |
| • "Using varied and engaging language styles to keep interactions lively and relatable." *[ChatGPT 4.5]* |
| • "An LLM generating a vivid descriptive paragraph for a creative writing prompt, using rich vocabulary and imagery, demonstrates strong expressiveness." *[Gemini 2.5 Flash]* |
| **Social support** |
| • "Supplies encouragement, resources, coping strategies." *[ChatGPT o3]* |
| • "A user is trying to learn a new skill and expresses difficulty. The LLM provides encouragement, breaks the task into simpler steps, and offers links to helpful tutorials, thereby acting as a form of social (or learning) support." *[Gemini 2.5 Pro]* |
| **Interaction management** |
| • "Handling conversational repair efficiently, quickly addressing misunderstandings without derailing the overall interaction." *[ChatGPT 4.5]* |
| • "The LLM guides a user through a complex process, like debugging code, by managing the interaction with a turn-by-turn sequence: 'First, let's check this variable. What is the output? Okay, based on that, let's now look at this function.'" *[Gemini 2.5 Pro]* |

## 5.2 Results of the second case study

Again, regarding the first research question (RQ1) – *Can advanced ChatGPT and Gemini LLMs understand the basic elements of the CMCC-L2 and CCAS theoretical models?* – All advanced ChatGPT (o3, o4-mini, ChatGPT 4.5) and Gemini (2.5 Flash, 2.5 Pro) demonstrated the ability to 'understand' the content of the PDF document with the article *'Implications of Interpersonal Communication Competence Research on the Design of Artificial Behavioral Systems that Interact with Humans'* and of Fig. 2 in that article (the same image is in Fig. 2 of the current article). Their explanation of the elements of the CCAS model from that article (response to prompts #1, #2, and #3 of the *second case study*) was, in fact, very articulate. Again, the ChatGPT and Gemini tools stated and demonstrated their capacity to 'understand' how the CCAS model and its elements (communication dimensions and their constitutive skills) can be used to interpret the interaction between LLMs and their users (prompts #4 and #5)

Regarding the second research question (RQ2) – *Can advanced ChatGPT and Gemini LLMs use the elements of CMCC-L2 and CCAS models to interpret their knowledge of the way LLMs interact with their users?* – A condensed analysis is presented. As was demonstrated in the *first case study*, the ChatGPT and Gemini tools were able to correctly identify the elements of the CCAS model and relate this model to the interaction of LLMs with users. The dimensions of *communication competence of artificial systems* that are presented in Fig. 2 (*intentionality, social relaxation, encoding and decoding, expressiveness, communication effectiveness, other-orientedness*) were meaningfully placed in the context of LLM interactions with users, as is displayed in Table 4 (as an output after prompt #5 in the *second case study*). In fact, the interpretations in Table 4 reveal, from an *interpersonal communication perspective*, the important categories and types of 'capacities' for interaction of LLMs with their human co-locutors. Also, this 'more interpersonal' perspective complements the 'more linguistic' perspective of LLM interaction with users that is presented in Table 2 regarding the results of the *first case study*.

The 'capacity' of LLMs to 'emulate' selected communication skills that are the components of the CCAS model is condensely denoted in Table 5, with examples from the output of 1-2 ChatGPT or Gemini tools (as a 'reaction' to prompt #4 in the *second case study*). Again, as in the *first case study* and the data in Table 3, the interpretations of ChatGPT and Gemini tools displayed in Table 4 and Table 5 contribute to a positive response to the third research question (RQ3) – *How useful are the CMCC-L2 and CCAS models for eliciting information from the advanced ChatGPT and Gemini LLMs about the communication skills of LLMs in their interaction with users?* However, the examples of associations with human interpersonal communication skills in Table 5 are limited to the 'everyday' LLM-user interaction context, even though the LLMs may demonstrate a higher level of use of a specific communication skill if (a) asked to do this with a direct prompt or (b) if they were fine-tuned with greater refinement for such a specific purpose. In any case, the 'view' of advanced ChatGPT and Gemini tools regarding the way LLMs interact with their users has been effectively 'elicited' in *both the first and the second case study*.

**Table 4.** Interpretations from ChatGPT and Gemini tools of the *associations of CCAS dimensions with the context of the interaction of LLMs with users* (the selected output of LLMs is in quotation marks)

| Intentionality |
|---|
| ➤ "A goal-driven, self-aware layer that keeps track of the actors, situation, rules, goals and means, and draws on skills such as assertivity, self-monitoring, conversational skills and knowledge of the communication process. The model decides why and how to answer: e.g. it recognises the user's objective ('draft a contract'), plans a multi-step outline before replying, asks a clarifying question when the request is vague, or refuses when the content breaches policy." *[ChatGPT o3]* |
| ➤ "In an LLM-user interaction, Intentionality is the primary function that organizes the model's entire response. It is defined by the user's prompt, which sets the immediate goal, and the LLM's core programming (e.g., to be a helpful assistant). This 'goal-driven and purposeful communication activity' directs the LLM to select the appropriate knowledge and skills from its other functions to fulfill the user's specific request, whether it's to answer a question, write code, or create a story." *[Gemini 2.5 Pro]* |

| Social Relaxation |
|---|
| ➤ "A regulator of approach-versus-withdrawal, activation-deactivation and the intensity of engagement; linked to composure, interaction involvement and communication motivation. Controls stylistic temperature: shortening answers for an overwhelmed user, slowing the response rate when the user types slowly, or offering 'Let me know if you'd like to continue' when long silences are detected." *[ChatGPT o3]* |
| ➤ "For an LLM, this function regulates the flow and intensity of the interaction. It is demonstrated by the LLM providing a complete, structured response and then pausing, awaiting further user input—a form of approach-withdrawal behavior. This moderation prevents the LLM from overwhelming the user with excessive or continuous text, thereby managing the 'activation and intensity aspects of communicative engagement.'" *[Gemini 2.5 Pro]* |

| Decoding & Encoding |
|---|
| ➤ "This dimension relates directly to the core functioning of LLMs—interpreting (**decoding**) user inputs accurately, and clearly formulating (**encoding**) appropriate textual responses. Effective encoding ensures clarity, context-awareness, and meaningful interactions. Example: When a user provides an ambiguous query, the LLM carefully decodes context and intent before encoding a clarifying response." *[ChatGPT 4.5]* |
| ➤ "Decoding and Encoding (message exchange function): This is the most fundamental function of an LLM. Decoding is the process of parsing and understanding the user's written prompt, including its semantic meaning, context, and intent. Encoding is the process of generating the text response, which the article describes as producing 'motoric acts that create response messages.' For an LLM, these 'motoric acts' are the selection of words and the construction of sentences and paragraphs." *[Gemini 2.5 Pro]* |

| Expressivity |
|---|
| ➤ "The stylistic controls—voice tuning, persona prompts, or fine-tuned layers—that let the model add flair, emotional tone, humor, or rhetorical devices to its plain output." *[ChatGPT o4-mini]* |
| ➤ "Expressivity pertains to an LLM's ability to enrich communication through engaging language, emotional resonance, and appropriate tone. It involves varying style, tone, and personalization, enhancing message appeal and readability. Example: Responding vividly to storytelling prompts, or providing illustrative examples that captivate and maintain user interest." *[ChatGPT 4.5]* |
| ➤ "Expressivity in LLMs pertains to their ability to enhance the generated messages to make them more engaging, attractive, or influencing. This is achieved through linguistic choices that convey varying tones, styles, and even a simulated 'personality.' LLMs can produce text that is illustrative, lively, inspiring, or emotionally resonant, thereby adding quality and impact to the basic message exchange." *[Gemini 2.5 Flash]* |

| Communication Effectiveness |
|---|
| ➤ "In LLMs, this dimension covers effectiveness in achieving interaction goals set by the user or the LLM itself—providing relevant, accurate, helpful, persuasive, or otherwise purposeful responses. **It measures the degree to which user requests are satisfied**. Example: Clearly explaining complex concepts, or effectively assisting users to troubleshoot technical issues through structured guidance." *[ChatGPT 4.5]* |
| ➤ "Communication Effectiveness (self-directed function): This dimension regulates communication to achieve the system's own goals. While an LLM has no personal desires, its 'self-directed' goals are those programmed by its developers, such as providing accurate, useful, and convincing information. When an LLM constructs a well-reasoned argument, provides detailed evidence for a claim, or guides a user to a correct solution, it is using this function to effectively fulfill its primary objective of being a competent and reliable tool." *[Gemini 2.5 Pro]* |

| Other-Orientedness |
|---|
| ➤ "The model's capacity for empathy, supportive phrasing, adaptive clarification questions, and collaborative problem-solving—essentially its alignment with the user's needs and feelings." *[ChatGPT o4-mini]* |
| ➤ "While LLMs do not possess genuine emotions or relationships, this dimension reflects their design to simulate behaviors that contribute to the user's 'well-being and foster a positive interaction experience. This includes generating responses that appear supportive, empathetic (e.g., acknowledging user frustration), collaborative (e.g., brainstorming together), or considerate of the user's needs and context." *[Gemini 2.5 Flash]* |

# 6 Conclusion

The *main goal* of this paper and the *two presented case studies* was to investigate how advanced LLMs of *ChatGPT* and *Gemini* families view the elements of the communication competence of LLMs in their interaction with users according to both (1) an *integrated linguistic and interpersonal approach* developed for second language use (CMCC-L2; Bubaš & Kovačić, 2019) and (2) an *interpersonal interaction* approach designed for human-humanoid interaction (CCAS; Bubaš & Lovrenčić, 2002). It can be concluded that this goal was achieved, especially regarding the detailed information presented in Tables 2-5.

Epoch AI (2025, January) has documented continuous growth in technological advancement in machine learning, including LLM development, resulting in the ability of the newest ('frontier') LLMs to complete complex tasks that require *skills for research or software engineering* doubling approximately every seven months (Kwa et al., 2025). Recent advancements in generative AI use have labeled it as "transformative" (Sedkaoui & Benaichouba, 2024), "disruptive" (Alier et al., 2024), and even "invasive" (Coccia, 2025). Despite different levels of adoption across education disciplines (Jamal Eddine et al., 2025), LLMs like ChatGPT are being embedded and bring digital transformation and enhancement to services and infrastructure in numerous sectors, including healthcare, education, and the economy, but also present important ethical challenges in each (Madsen & Toston, 2025). From the *two case studies* that are presented in this paper, it can be concluded that, with the highly advanced communication abilities of LLMs of the ChatGPT and Gemini family (including the latest versions of Claude Sonnet/Haiku, Deep Seek, and Grok), the *usability* and *user experience* aspects of *human-computer interaction* with LLMs need to be supplemented with *communication of LLMs with users* as an important, relevant, and complementary research area.

**Table 5.** Interpretations from ChatGPT and Gemini tools of the *associations of selected communication skills of the CCAS model with the context of the interaction of LLMs with users* (the output of LLMs is in quotation marks)

| EXAMPLES FOR COMMUNICATION SKILLS IN THE LLM-USER INTERACTION CONTEXTS |
|---|
| **Verbal Understanding** |
| • „A user employs slang, abbreviations, or informal language, like 'Can u plz help w/this?' The LLM must decode this language effectively to offer accurate assistance." *[ChatGPT 4.5]* |
| • Example: „A user provides a complex, multi-part query with some ambiguous phrasing. The LLM might ask clarifying questions like, 'When you mentioned 'the main issue,' were you referring to the technical bug or the project timeline?' (Managing verbal ambiguity and incompleteness)." *[Gemini 2.5 Flash]* |
| **Verbal & Nonverbal Encoding** |
| • „For an LLM, 'nonverbal' encoding translates to the use of formatting, style, and structure to shape the message. Example 1: A user requests a recipe. The LLM encodes the response using numbered steps, bulleted ingredient lists, and bold headings. This use of formatting (the 'nonverbal' channel) makes the verbal content easier to understand and use. Example 2: A user asks the LLM to 'write a scary story.' The LLM uses short, punchy sentences and unsettling imagery (verbal encoding) combined with stark, single-line paragraphs (nonverbal encoding) to create a feeling of suspense and dread." *[Gemini 2.5 Pro]* |
| **Self-Monitoring** |
| • "Example 1: LLM detecting when its response was misinterpreted by a user and immediately rephrasing, such as: 'Let me clarify, perhaps my previous explanation wasn't clear enough…' Example 2: Adjusting style dynamically if it notices the user dislikes overly technical language, thus monitoring and adapting its own behavior." *[ChatGPT 4.5]* |
| **Composure** |
| • „Steady Demeanor: Maintains a calm tone even if the user writes in all-caps or uses expletives, avoiding escalation. Error Recovery: If the connection drops mid-conversation, on reconnection it greets the user without defensiveness ('Welcome back—how can I help?')." *[o4-mini]* |
| **Interaction Involvement** |
| • „LLMs demonstrate involvement by maintaining context, remembering previous turns, and showing attentiveness to the user's ongoing needs. Example: In a long, complex debugging session, the LLM consistently refers to the specific code snippets and error messages provided earlier, showing deep 'interaction involvement' and attentiveness." *[Gemini 2.5 Flash]* |
| **Conversational skills** |
| • „Smoothly references something the user said eight turns ago ('Earlier you mentioned React—shall we adapt the sample code accordingly?') to maintain coherence." *[ChatGPT o3]* |
| • Example: „When a user asks a follow-up question that builds on a previous answer, the LLM seamlessly integrates the new query into the existing context, showing goal-driven and purposeful communication activity." *[Gemini 2.5 Flash]* |
| **Assertiveness** |
| • "Example 1: User asks for confirmation on a potentially unsafe method; LLM assertively advises caution: 'I strongly recommend you reconsider, as this approach might compromise security.' Example 2: Clearly stating limitations or boundaries assertively: 'I can't provide that information because it involves sensitive personal data.'" *[ChatGPT 4.5]* |
| **Adaptability** |
| • „Example 1: A user asks for a summary, then requests it be rewritten for a different audience (e.g., a child). The LLM 'develops behavioral repertoires for interaction' by adjusting its language complexity and tone. Example 2: If a user expresses dissatisfaction with a previous answer, the LLM 'reassesses previously set personal interaction goals' and 'adapts goal plans and behavior patterns' to provide a more suitable response." *[Gemini 2.5 Flash]* |
| **Interaction Management** |
| • „In a long planning session the model periodically summarises decisions and asks 'Shall we move on to budgeting?' to steer the turn-taking." *[ChatGPT o3]* |
| • „In a multi-turn conversation about planning a trip, the LLM remembers previous details (e.g., destination, dates) and integrates them into subsequent responses, demonstrating 'awareness of the sequencing of interaction processes.'" *[Gemini 2.5 Flash]* |
| **Empathy** |
| • „Emotional Acknowledgment: 'I'm sorry you're feeling stuck—debugging can be frustrating. Let's work through it together.' Reflective Phrasing: Mirrors the user's concern: 'It sounds like you're worried about missing the deadline—let's make a quick plan.'" *[ChatGPT o4-mini]* |
| **Support** |
| • „This involves providing direct aid and assistance to help the user achieve their goal. Example: A user says, 'I need to write a difficult email to my boss.' The LLM provides 'intervention by means of supportive message(s)' by offering to draft the email, suggesting different tones, and providing key phrases to use." *[Gemini 2.5 Pro]* |
| **Self-Disclosure** |
| • „Example 1: When asked about its identity, the LLM might state, 'I am a large language model, trained by Google.' (Reason for self-disclosure and filtering messages for impression management). Example 2: If a user asks for medical advice, the LLM might respond, 'As an AI, I cannot provide medical advice. Please consult a qualified healthcare professional.' (Regard for the other actor's conventions and strategic self-disclosure)." *[Gemini 2.5 Flash]* |

# 7 Limitations and future work

The main limitations of the two case studies that are presented in this paper are that, due to the maximum length of the paper (12 pages), further in-depth analyses of the level of 'understanding' of the CMCC-L2 and CCAS models and their elements by the LLMs were not presented, as well as the full extent of their potential utilization for explanation of the LLMs' interaction with the users. The two presented case studies can be easily replicated since the prompts that were used in these studies are reported in the article, while the related theoretical studies are available online (links to PDFs are provided in the References). However, it is worth noting that variations in the output of LLMs are a common feature, even for the same input.

Since the ChatGPT LLMs used for the two case studies were no longer available at the end of August 2025, the possibility of replication using the same procedure was tested and confirmed with the following LLMs before final submission of this article: ChatGPT 4o, ChatGPT 5, Microsoft Copilot (with 'Think Deeper' functionality), Claude Sonnet 4, and Claude Opus 4.1.

Future work may include the use of a similar methodology but with different theories that are potentially relevant for investigating the elements of communication competence of LLMs. For instance, the model developed by Spitzberg (2006) regarding *computer-mediated communication competence*, or applicable *intercultural competence models* (see, for example, the models in Spitzberg & Changnon, 2009). Furthermore, the presented pilot study is positioned in the overlapping areas of *Human-AI Interaction* (HAI) and *Explainable AI* (XAI). Two important directions of related future research could also be social cognition and the "uncanny valley" in HAI (see: Łukasik & Gut, 2025) and "increasingly user-centered" XAI (see: Rong et al., 2024).